\RequirePackage{fix-cm}
\documentclass[twocolumn,epjc3]{svjour3}
\smartqed  
\RequirePackage{graphicx}
\usepackage{tikz}
\usetikzlibrary{patterns}
\usepackage{axodraw}
\usepackage{cite}
\usepackage{epsfig,multicol}
\usepackage{lineno}
\usepackage{amsmath,amssymb,mathrsfs}

\newcommand\lsim{\mathrel{\raise.3ex\hbox{$<$\kern-.75em\lower1ex\hbox{$\sim$}}}}
\newcommand\gsim{\mathrel{\raise.3ex\hbox{$>$\kern-.75em\lower1ex\hbox{$\sim$}}}}

\newenvironment{Eqnarray}%
     {\arraycolsep 0.14em\begin{eqnarray}}{\end{eqnarray}}
\newcommand{\beqa}{\begin{Eqnarray}}
\newcommand{\eeqa}{\end{Eqnarray}}
\newcommand{\beq}{\begin{equation}}
\newcommand{\eeq}{\end{equation}}

\journalname{Eur. Phys. J. C}
\begin{document}
\title{
Exclusion of heavy, broad resonances from precise measurements of $WZ$ and $VH$ final states at the LHC
%
}
%
\titlerunning {Exclusion of heavy, broad resonances from precise measurements of $WZ$ and $VH$ final states at the LHC}
\author{You-Ying Li\thanksref{addr1}
  \and
  Rosy Nicolaidou\thanksref{addr2}
  \and
  Stathes Paganis\thanksref{e3,addr1}
}
%
\thankstext{e3}{stathes.paganis@cern.ch, corresponding author}
\authorrunning{You-Ying Li, R.Nicolaidou, S.Paganis}
\institute{
 Department of Physics, National Taiwan University, 
No 1, Sec 4, Roosevelt Road, Taipei 10617, Taiwan\label{addr1}
 \and ~IRFU, CEA, Universite Paris-Saclay, Gif-sur-Yvette; France\label{addr2}
}
\date{Received: date / Accepted: date}

\maketitle

\begin{abstract}
A novel search for heavy vector resonances 
in the $H\rightarrow b\bar{b}$ and $Z\rightarrow b\bar{b}$ final states
in association with a leptonically decaying $V$ ($Z$ or $W$)
and $W$-only respectively, is proposed.
It is argued that excesses with respect to the Standard Model prediction 
should be observed in all final
states (0, 1 or 2 leptons), with the 1-lepton final state being the strongest.
Since the relative strengths of these excesses depend on branching ratios
and efficiencies, this is a clear signal for the presence of heavy resonances or their
low mass tails. A general vector-triplet model is used to explore
the discovery potential as a function of the resonance mass
and width. Recent Higgs to $b\bar{b}$ observation data 
reported by the experiments ATLAS and CMS are used to test the model. Current limits 
are extended to resonance widths over mass as large as 9\%.
\keywords{Computational methods and analysis tools, Hadron and lepton
  collider physics}
\PACS{
      {29.85.Fj}{Data analysis} \and
      {14.80.Bn}{Standard-model Higgs bosons}
     } 
\end{abstract} 
%

\section{Introduction}
\label{intro}
Heavy vector resonances naturally appear in several extensions of the Standard
Model (SM), such as GUT theories \cite{Pati,Georgi,Fritzcsh}, composite Higgs
\cite{Eichten,Contino}, little Higgs \cite{lHiggs1,lHiggs2,lHiggs3}, and models
with vector $Z^\prime$ \cite{Zprime1, Zprime2}, and $W^\prime$ models \cite{Wprime}.
The LHC experiments, in most cases, have
performed direct searches for heavy narrow resonances decaying to
dibosons and $VH$ final states and have put limits to masses up to
about 5.5~TeV
\cite{ATLASVV_jjjj_19Dec17,
ATLASVV_02lep_19Aug17,
ATLASVV_1lep_19Oct17,
ATLASVHbb17Dec17,
ATLASVHbb_jj_21Jul17,
CMSVHbb_leptons_8Jul18,
CMSVV_llqq_6Dec17,
CMSVV_lvqq_12Dec17,
CMSVV_vvqq_11Jul17,
CMS_VH_17,
CMSVVandVH_17,
ATLASHeavyCombination}.

Broader resonances ($\Gamma/m > 4\%$) appear for larger values 
of the resonance coupling strength to weak bosons and the Higgs, $g_V$,
and depending on their mass, width, and production cross
section, the LHC experiments can be sensitive to a large part of the
resonance distribution. It is interesting to study to what extent the
experiments can be sensitive to the full distribution or
tails of such resonances and what the best observables are.
In this work, we start addressing this question with
a specific final state: a pair of $b$ quarks produced in association
with a weak gauge boson ($Z$ or $W$) that subsequently decays
leptonically (where leptons $\ell$ are electrons and muons).
The interest in this final state is because in the large coupling regime,
$g_V>3$, the heavy boson branching fraction is dominated by decays to $WZ$ and $VH$, 
while the decays to fermions are suppressed \cite{ourCHpaper1}.
At the same time, both the Drell-Yan $q\bar{q}$ and the VBF production
modes contribute, providing non-negligible cross sections.
Here, we propose as an early indirect signal of the presence of vector
resonances, the simultaneous excess in the 0, 1, and 2-lepton
final states, for both
$WZ\rightarrow \ell\nu b\bar{b}$ and $VH\rightarrow \ell\ell(\nu) b\bar{b}$, in a measurement
of the $b\bar{b}$ quark invariant mass $M_{bb}$. 
The strength of these excesses in the different final states
can be predicted by branching ratios and experimental efficiencies.
We argue that the
0, 1, and 2-lepton final states should all be
sensitive, albeit with very different sensitivity,
to exotic $V^\prime\rightarrow VH$ decays. In particular, the
1-lepton state will be the most sensitive in Higgs excesses
and in addition, it should also show a smaller level of excess in the $Z$-boson
$b\bar{b}$ peak. The pattern of these excesses should be a clear sign of a
heavy vector triplet (HVT), even before its observation as a (broad)
mass resonance at the TeV scale. Using a general HVT framework that is
also used by the experiments to model 
such heavy vector triplets \cite{hvt1}, we derive the sensitivity of
a single LHC experiment as a function of the resonance width.

In this paper, we first introduce the HVT framework and 
summarize the recent LHC results. In Section 3 we present
an analysis used to search for heavy vector triplets at the LHC, and we
report on the search potential for the $b\bar{b}+$0,~1, and 2-lepton  final 
states, for a range of masses and widths.
Finally, the compatibility of our predictions with the recently
published 13 TeV $H\rightarrow b\bar{b}$ data from ATLAS and CMS 
in terms of expected and observed limits, is discussed.

\section{BSM Heavy Vector Resonances}
\label{hvt}
A general HVT phenomenological Lagrangian can be used for the modelling
of resonances predicted by a wide range of Beyond the Standard Model (BSM) scenarios \cite{hvt1}. 
The Lagrangian describing the interactions of these resonances $V^{a \prime}$, $a=1,2,3$ with quarks, 
leptons, vector bosons and the Higgs boson is shown below:
\begin{eqnarray}
\nonumber
\mathcal{L}_{V}^{int}=&-&\frac{g^2c_F}{g_V}V_{\mu}^{a \prime}\bar{q}_k\gamma^{\mu}\frac{\sigma_{a}}{2}q_{k}
-\frac{g^2c_F}{g_V}V_{\mu}^{a \prime}\bar{\ell}_k\gamma^{\mu}\frac{\sigma_{a}}{2}\ell_{k}\\\nonumber
&-&g_Vc_H\left(V_{\mu}^{a \prime}H^{\dagger}\frac{\sigma^a}{2}iD^{\mu}H+\mathrm{hc}\right),\nonumber
\end{eqnarray}
where $q_k$ and $\ell_k$ are the quark and lepton weak doublets, $H$ is the Higgs 
doublet and $\sigma^a$ the three Pauli matrices.
In this Lagrangian, the HVT triplet
$V^{a \prime}=(W^{+\prime},W^{-\prime},Z^{\prime})$ interacts with the Higgs
doublet, i.e. the longitudinal degrees of freedom of the SM $W$ and
$Z$ bosons and the SM Higgs, with a coupling strength $g_V$. In
order to allow for a broader class of models, this coupling strength
can be varied by the parameter $c_H$, so in the Lagrangian the
full coupling to the SM weak and Higgs bosons is $g_Vc_H$.
The HVT resonances also couple to the SM fermions, again
through their coupling to the SM weak and Higgs bosons,
$g^2/g_V$, where $g$ is the SM $SU(2)_L$ weak gauge
coupling. This coupling between HVT resonances and fermions is also
controlled by an additional parameter $c_F$ to allow for a broader
range of models to be included, as follows: $g^2c_F/g_V$.
As discussed in \cite{ATLASHeavyCombination}, the experiments consider two Drell Yan (DY) production
scenarios: model A is a scenario that reproduces 
the phenomenology of weakly coupled models based on an extended
gauge symmetry \cite{Zprime1}. 
In this case, the couplings are $\frac{g^2c_F}{g_V}=-0.55$ and
$g_Vc_H=-0.56$, with the fermion coupling being universal. 
The second DY scenario, referred to as model B, implements a strongly
coupled scenario as in composite Higgs models with $\frac{g^2c_F}{g_V}=0.14$ and  $g_Vc_H=-2.9$. 
In model B, the $V^{\prime}$ resonances are broader than 
in the weakly coupled scenario, model A, but for $|g_Vc_H|\leq 3$ they remain narrow relative to
the experimental resolution. For $|g_Vc_H|>3$, the resonance intrinsic 
width becomes significant and cannot be neglected.
In summary, the ATLAS and CMS benchmarks correspond 
for Model A to $c_H=-\frac{g^2}{g_{V}^2}$, $c_F=-\frac{1}{3}$ and
$g_V=1$, and for model B to $c_H=-1$, $c_F=1$ and $g_V=3$.
ATLAS and CMS experimental data from direct $m_{VH}$ and $m_{VV}$
searches, exclude part of the
parameter phase space $(g_V, c_H, c_F)$, for which the intrinsic
width $\Gamma$ of the new bosons is dominated by the experimental
resolution ($\sim 4\%$ of the mass).
This is also the case, for example, for 
$|g_Vc_H|\leq 3$, where negligible natural width is assumed (narrow width assumption).
The goal of this work is to explore the part of the HVT parameter 
space with $|g_Vc_H|>3$, where the heavy resonances have a significant 
natural width. Following the model B benchmark, we fix the two constant factors
to $c_H=-1$, $c_F=1$ and allow $g_V$ to vary.

\begin{figure}[hbt!]
\begin{center}
\resizebox{0.5\textwidth}{!}{
\includegraphics{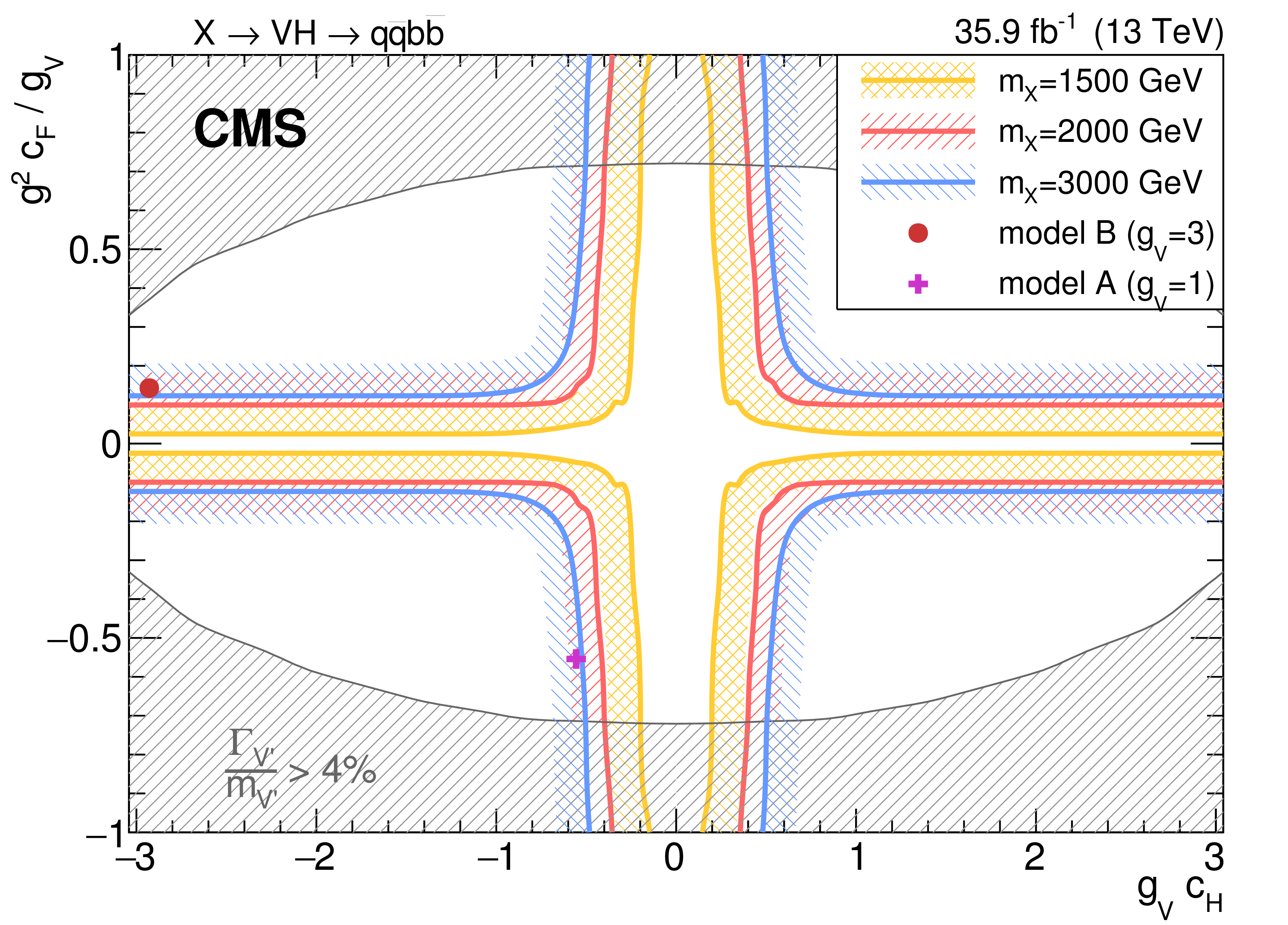}
}
\caption{
CMS, \cite{CMS_VH_17},
observed 95\% CL exclusion contours in the HVT parameter space
${g_Vc_H, (g^2/g_V)c_F}$ for narrow resonances of mass 1.5 TeV, 2.0 TeV and
3.0 TeV. Due to the narrow width approximation assumption,
the exclusion validity is restricted to roughly $|g_Vc_H|\leq 3$ and outside the 
hatched area shown,
thus leaving a large part of the available
phase space unconstrained ($3<|g_Vc_H|<4\pi$).
Similar limits have been published by ATLAS, \cite{ATLASVHbb17Dec17}. 
}
\label{HVTnarrowATLAS}
\end{center}
\vspace{-5mm}
\end{figure}

Results from recent experimental analyses that exclude part
of the HVT phase space are shown in Fig.~\ref{HVTnarrowATLAS}.
The validity of these exclusions is actually limited due to the narrow width approximation. 
For larger, $|g_Vc_H|>3$, couplings, the width grows and becomes comparable
to or larger than
the experimental resolution, 
leaving a large part of the HVT phase space unexcluded: 
$3<|g_Vc_H|<4\pi$, where $g_V\simeq 4\pi$ is the perturbative limit. 
This is true even for resonance masses of order TeV.

In this work we argue that the presence of
non-zero width resonances, could be observed 
as an excess of events in the $VH\rightarrow \ell\ell(\nu)b\bar{b}$ and
$WZ\rightarrow \ell\nu b\bar{b}$ decays with 0, 1, or 2 leptons
in the final state. Independent of the fact that due to the experimental resolution
the close-by $Z\rightarrow b\bar{b}$ and $H\rightarrow b\bar{b}$ peaks
have a partial overlap, in the 1-lepton case the excess
must be present in both peaks, while in the 0 and 2-lepton final
states it should appear only in the Higgs peak ($Z^{\prime}$ does not couple to $ZZ$). In addition, the 
excess in the 1-lepton category should be more significant, since 
in this case both $W^+$ and $W^-$ contribute.
A combined
analysis of the three final states can quantify the excess
and correlate it to events with $b\bar{b}$ pairs of higher $p_T$
than in the SM Higgs production, since
these pairs originate from heavy TeV-scale objects.

\begin{figure}[hbt!]
\begin{center}
\resizebox{0.4\textwidth}{!}{
\includegraphics{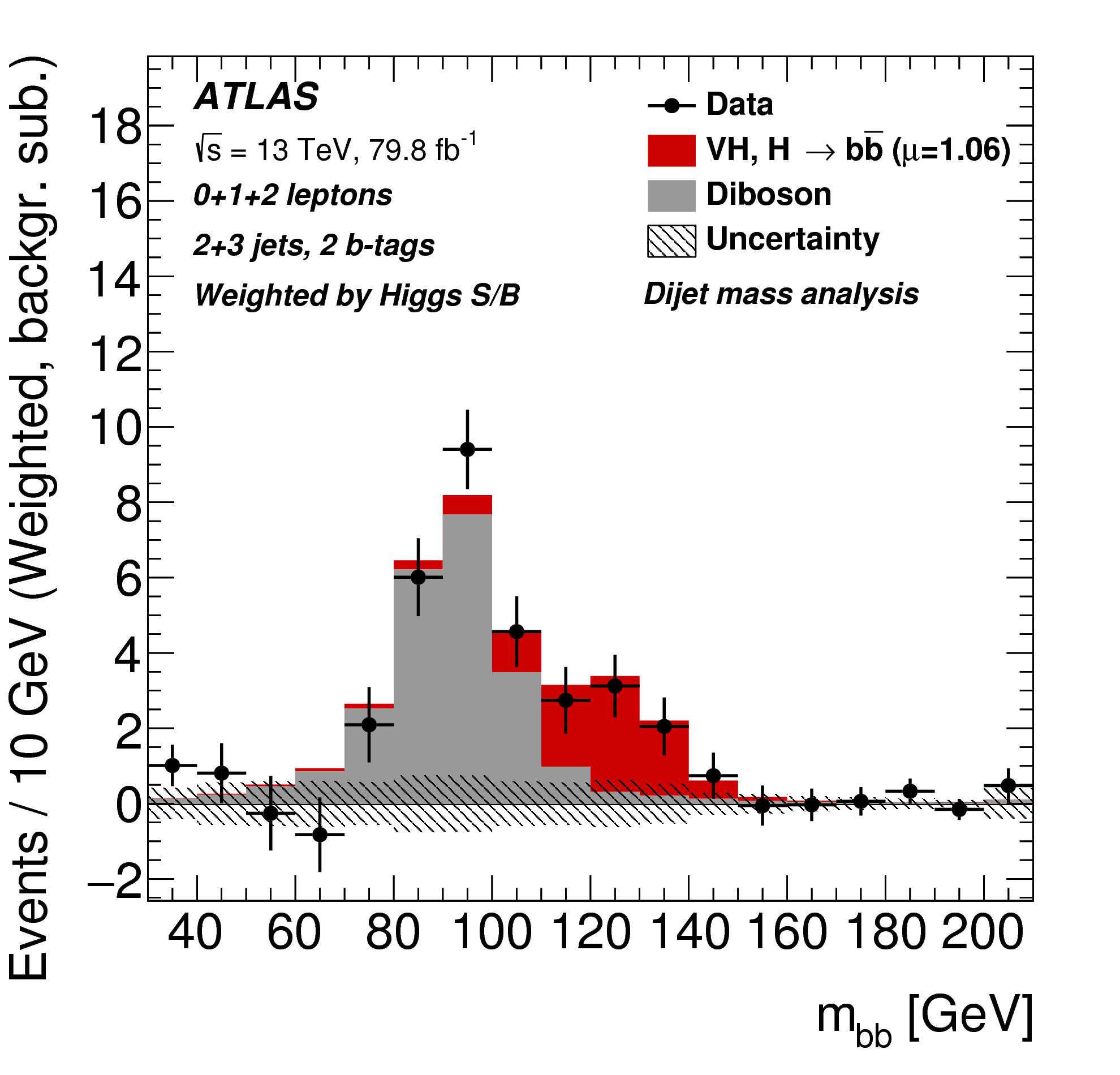}
}
\caption{Weighted bottom anti-bottom quark invariant mass data from
  ATLAS for 79.8~fb$^{-1}$\cite{ATLASSMvh}. The plot includes all
  final states with 0, 1 and 2 leptons and it compares the data yield
  with 1.06$\times$ the SM expectation of $H\rightarrow b\bar{b}$, 
  and the SM expectation of $Z\rightarrow b\bar{b}$.
}
\label{MbbATLAS}
\end{center}
\vspace{-5mm}
\end{figure}

\begin{figure}[hbt!]
\begin{center}
\resizebox{0.4\textwidth}{!}{
\includegraphics{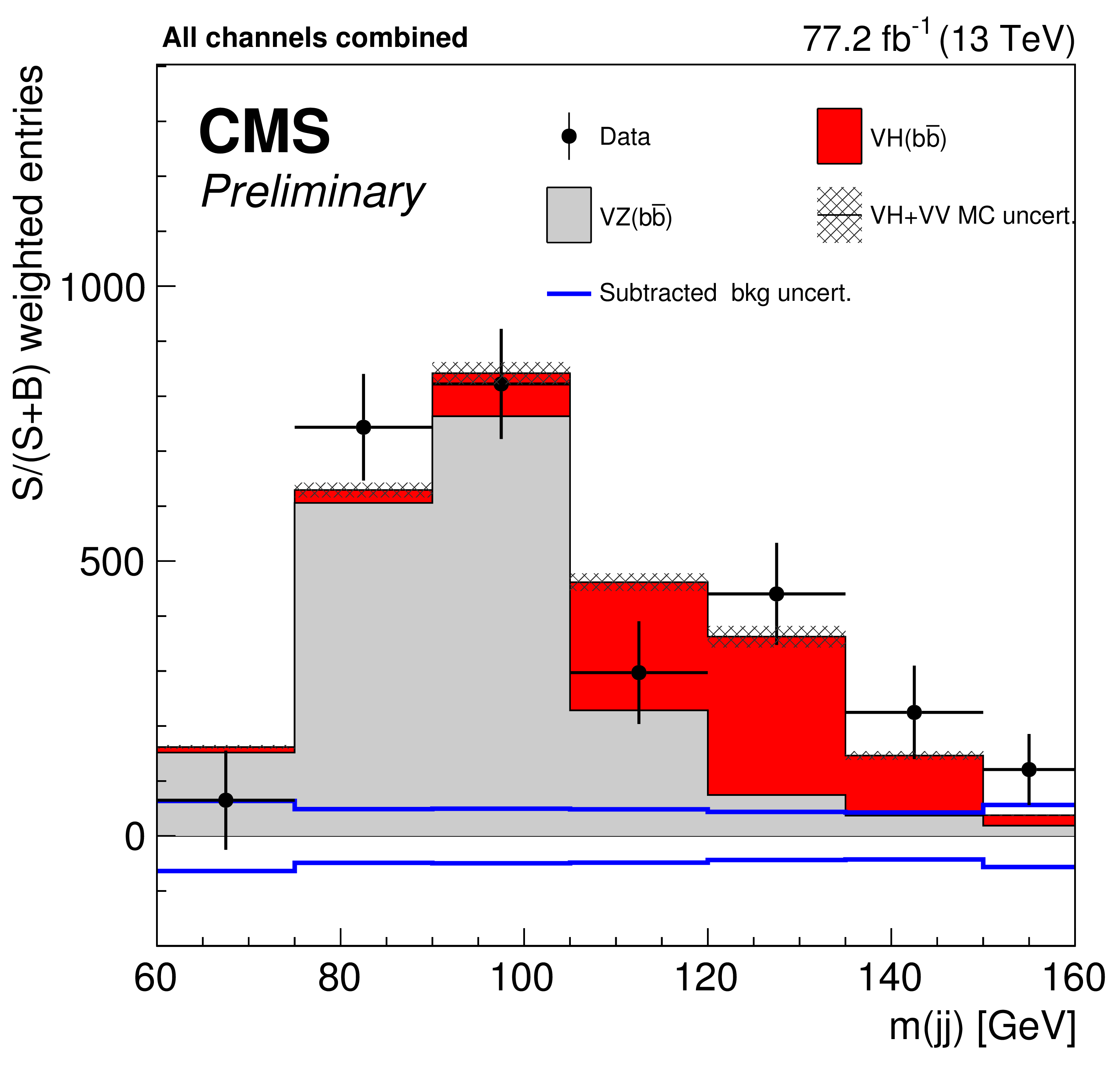}
}
\caption{Weighted bottom anti-bottom quark invariant mass data from
  CMS for $\sim 77.2$~fb$^{-1}$\cite{CMSSMvh}. 
The plot includes all final states with 0, 1 and 2 leptons and it compares the data yield 
with the SM expectation of $H\rightarrow b\bar{b}$ and the SM expectation of $Z\rightarrow b\bar{b}$.
}
\label{MbbCMS}
\end{center}
\vspace{-5mm}
\end{figure}

ATLAS and CMS recently published results in the search for the SM
$H\rightarrow b\bar{b}$ in association with a gauge boson decaying
leptonically \cite{ATLASSMvh,CMSSMvh}. This final state is identical
with the one we propose here having sensitivity to HVT resonances.
Both experiments report the measured $M_{bb}$ invariant
mass as shown in Figures~\ref{MbbATLAS} and \ref{MbbCMS}, 
for an integrated luminosity of $\sim 80$~fb$^{-1}$ at $13$~TeV. 

\begin{figure}[hbt!]
\begin{center}
\resizebox{0.5\textwidth}{!}{
\includegraphics{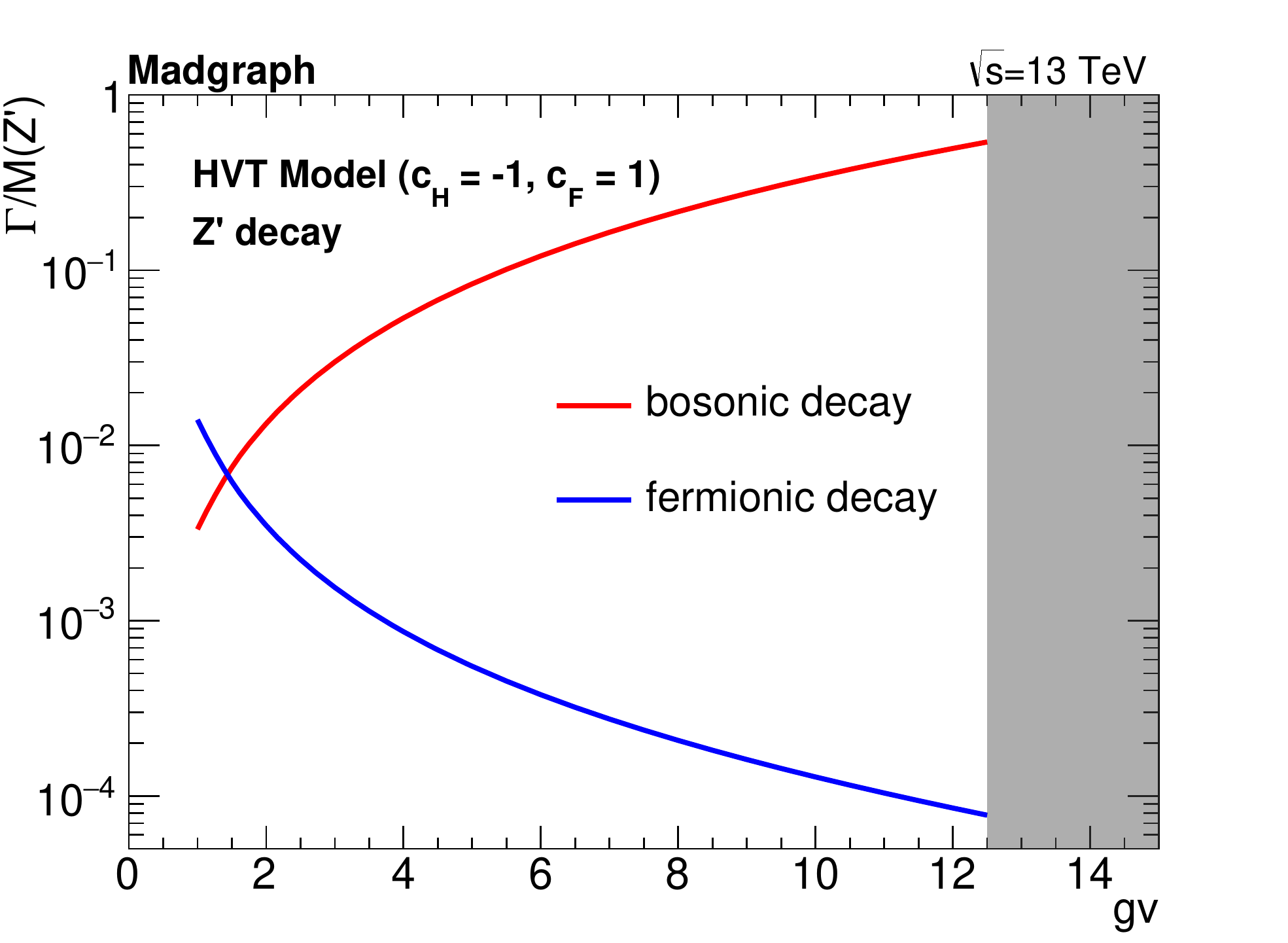}
}
\caption{
Heavy resonance width for bosonic and fermionic decays 
as a function of the $g_V$ coupling.
}
\label{WidthVsgV}
\end{center}
\vspace{-5mm}
\end{figure}

\begin{figure}[hbt!]
\begin{center}
\resizebox{0.5\textwidth}{!}{
\includegraphics{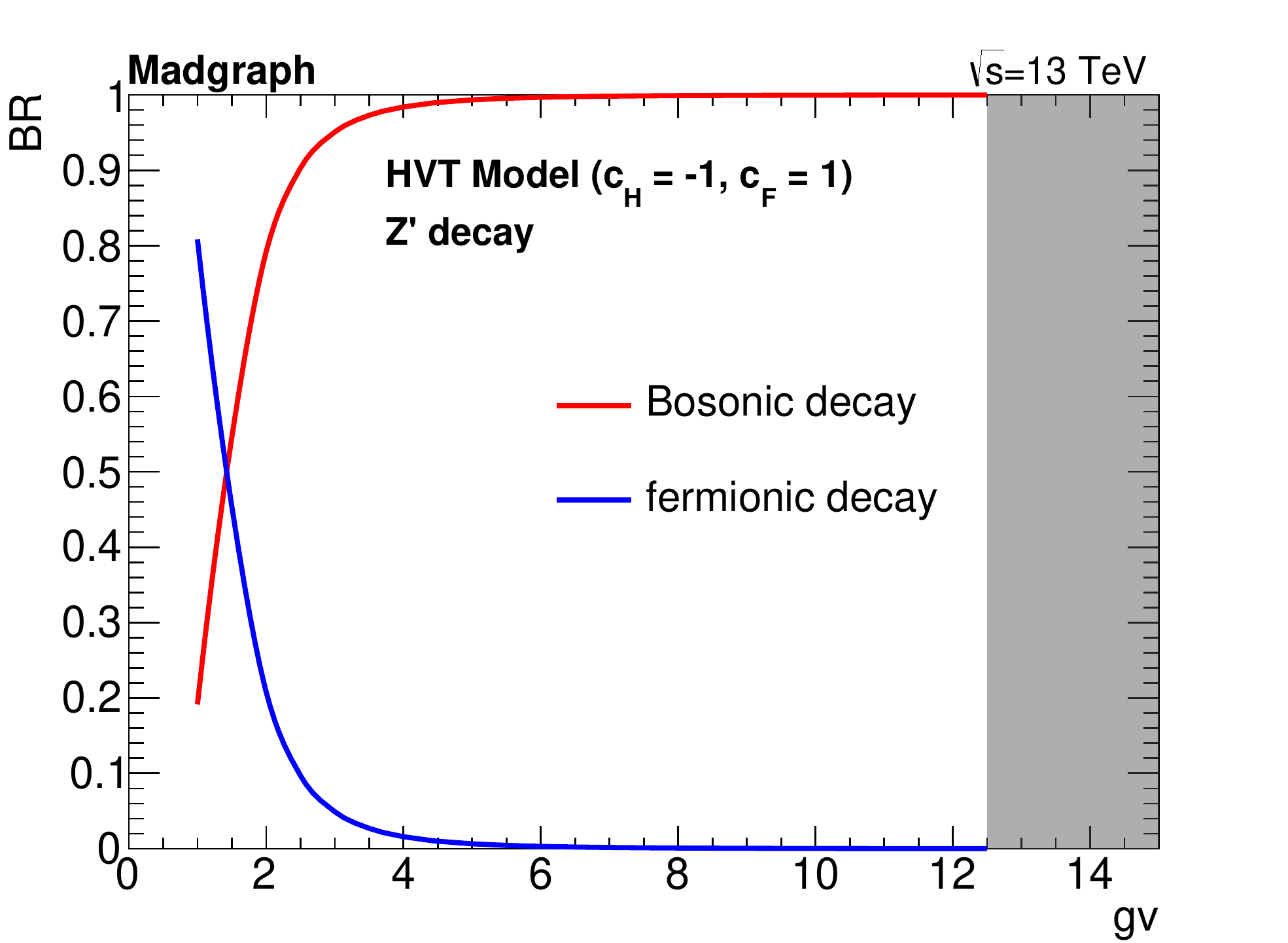}
}
\caption{Heavy resonance branching ratio to bosonic and fermionic
  decays as a function of the $g_V$ coupling.
}
\label{BRVsgV}
\end{center}
\vspace{-5mm}
\end{figure}

Broad heavy resonances mainly decaying to dibosons and pairs of fermions,
occur for larger values of the $g_V$ coupling ($g_V > 3$ for $c_H=-1$ and
$c_F=1$).
The resonance width as a function of
$g_V$ is shown in Fig.~\ref{WidthVsgV} and the branching ratios to a
pair of fermions and
diboson final states are shown in Fig.~\ref{BRVsgV}.
%
In both figures the calculations shown were performed by 
implementing the HVT model in MadGraph5~\cite{madgraph5}.
Fig.~\ref{WidthVsgV} shows that for $g_V > 3$ the resonance
natural width over the mass ratio exceeds $\sim 3\%$,
and as it can be seen in Fig.~\ref{BRVsgV}, the resonance decay 
to dibosons ($VH$, $VV$) is dominant. 
Direct searches for resonances in the $VV$ and $VH$ channels assume
narrow resonance width that corresponds to $g_V\leq 3$, leaving 
unexcluded a large part of the HVT model parameter space.

\begin{figure}[hbt!]
\begin{center}
\resizebox{0.48\textwidth}{!}{
\includegraphics{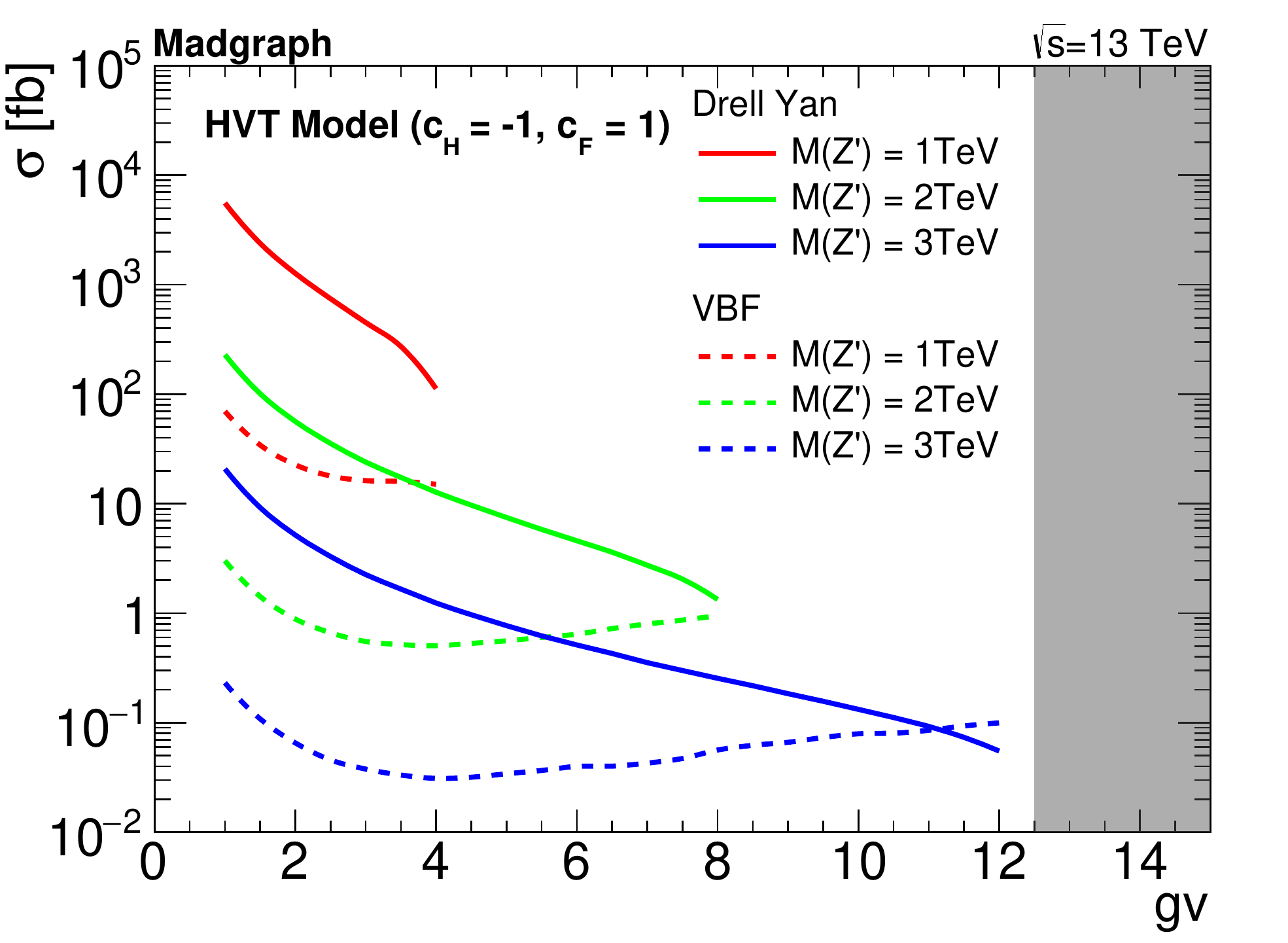}
}
\caption{Heavy resonance cross sections for the DY and VBF production modes, as a function of the
$g_V$ coupling.
}
\label{xsectionsVsgV}
\end{center}
\vspace{-5mm}
\end{figure}

The magnitude of the impact of these resonances 
on SM measurements depends on their
cross section which in turn, depends on the mass and the $g_V$ coupling.
Heavy resonance Drell-Yan and Vector Boson Fusion (VBF)
production cross sections for masses in the range 1-3~TeV are shown in Fig.~\ref{xsectionsVsgV}.
The DY mode is always dominant up to large values of $g_V$ close to the perturbative limit
$g_V\simeq 4\pi$, and the total cross section drops with $g_V$.
In Fig.~\ref{xsectionsVsgV} we can see that in the mass range
$1-2$~TeV, the cross sections for $g_V > 3$ range from $O(1)$ to
$O(100)$~fb, leading to non-vanishing contributions at present and
future measurements.
For parts of the parameter space corresponding to low resonance masses and high $g_V$
that are theoretically excluded (parts for which the input electroweak
parameters $\alpha_{EW}$, $G_F$ and $M_{Z}$,
are not reproduced by the HVT model), 
the cross section is not provided in Fig.~\ref{xsectionsVsgV}.

%
%
\section{Resonance Search Potential at Colliders}
\label{analysis}
To evaluate the expected signal yields for a broad
range of masses and widths, we rely on Monte Carlo
samples generated with MadGraph5, \cite{madgraph5}, interfaced to
PYTHIA \cite{Pythia8:2008}. To simulate the response of an
LHC-like experiment, realistic resolution and reconstruction
efficiencies for electrons, muons, photons and jets were applied with
the Delphes framework \cite{delphes}. The specific cut-based event
selection was based on the recent $VH$ with
$H\rightarrow b\bar{b}$ results from ATLAS and CMS \cite{ATLASSMvh,CMSSMvh}.
As discussed later, the efficiencies were normalized to the experimental ones. 
The selected events were split
in three categories according to the final state: two $b$ jets with 0, 1
or 2 leptons. We follow more closely the ATLAS selection including
the requirement of 2 or more jets out of which exactly two must be
$b$ jets. 
We will call this SM selection, the baseline selection, 
in order to differentiate from additional
discriminants which enhance the BSM Higgs signal.
The BSM Higgs signal decays to a $b\bar{b}$ pair 
with a transverse momentum, $p_{\perp}$,
significantly larger than the SM Higgs.
For this reason, an additional requirement applied to the baseline analysis,
is a cut on the transverse momentum of the $b\bar{b}$ system.
As in the LHC analyses, we only
consider resolved $b\bar{b}$ pairs although the search can be extended
to a single, fat $b\bar{b}$ jet analysis. As a discriminant, the $b\bar{b}$ invariant mass
$M_{bb}$ is used, which after full selection and subtraction of the $b\bar{b}$
continuum shows two peaks due to the presence of
$Z\rightarrow b\bar{b}$ and $H\rightarrow b\bar{b}$.

\begin{figure}[hbt!]
\begin{center}
\resizebox{0.45\textwidth}{!}{
\includegraphics{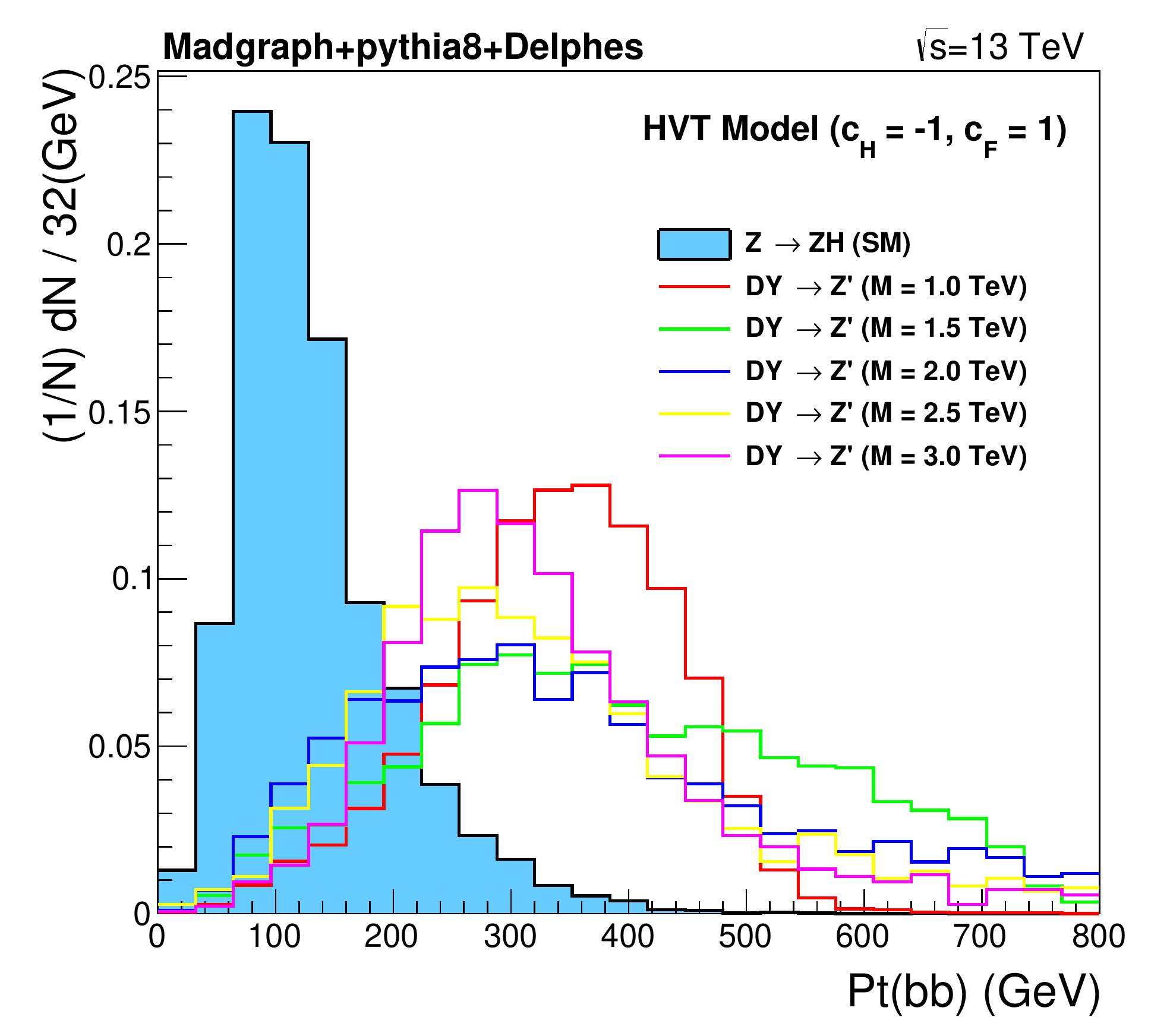}
}
\caption{Exotic Higgs$\rightarrow b\bar{b}$ transverse momentum, $p_{\perp}$, originating
  from heavy vector resonances, compared with SM Higgs $p_{\perp}$.
  Exotic Higgs bosons are typically of higher $p_{\perp}$ ($p_{\perp}>150$~GeV).
}
\label{ptbb}
\end{center}
\vspace{-5mm}
\end{figure}

The characteristic of the BSM signal is that the non-SM produced
$H\rightarrow b\bar{b}$ has high $p_{\perp}$. This can be seen
in Fig.~\ref{ptbb}, where the exotic Higgs has typical
$p_{\perp}>150$~GeV while the SM Higgs is softer. 
%
The expected signal and resonant SM background yields for the baseline
selection are presented in Table~\ref{yields1} for the 0, 1 and
2-lepton final states, for a resonance of mass 1.5~TeV, $g_V=5$ and an integrated
luminosity of 100~fb$^{-1}$.
For a $p_{\perp}>200$~GeV cut, we observe a significant increase in the fraction
of BSM signal over the SM resonant yields, that allows to estimate the
discovery potential as a function of the HVT model parameters.
\begin{table}
\small
  \begin{center}
    \begin{tabular}{cccccc}\hline
      Process           & $\sigma\times BR$ & A$\times \epsilon$ & Yield & Yield\\
      $qq\rightarrow V^\prime$ &  [fb]  & [\%]           & [$100$fb$^{-1}$] & $p_{\perp}>$\\
                        &        &                &           & $200$~GeV \\
\hline
$Z^{\prime}\rightarrow ZH\rightarrow \nu\nu b\bar{b} $ & 1.58 & 2.45 & 3.78 & 3.51 \\
$Z\rightarrow ZH\rightarrow \nu\nu b\bar{b} $ & 97.2 & 5.01 & 486 & 224 \\
$ZZ \rightarrow \nu\nu jj $ & 2580 & 0.27 & 697 & 248 \\
\hline
$W^{\prime}\rightarrow WH\rightarrow \ell\nu b\bar{b} $ & 3.87 & 5.5 & 21.3 & 19.4 \\
$W^{\prime}\rightarrow WZ\rightarrow \ell\nu jj $ & 3.79 & 0.23 & 0.81 & 0.59 \\
$W\rightarrow WH\rightarrow \ell\nu b\bar{b} $ & 225 & 1.44 & 324 & 148 \\
$WZ\rightarrow \ell\nu jj $ & 4148 & 0.13 & 529 & 173  \\
\hline
$Z^{\prime}\rightarrow ZH\rightarrow \ell\ell b\bar{b} $ & 0.553 & 4.76 & 2.6 & 2.2  \\
$Z\rightarrow ZH\rightarrow \ell\ell b\bar{b} $ & 34.2 & 13.7 & 467 & 68.6 \\
$ZZ\rightarrow \ell\ell jj $ & 910 & 0.96 & 875 & 72.6 \\
    \hline
    \end{tabular}
    \caption{Signal yield and resonant SM background yields for a
      1.5~TeV resonance, $g_V=5$ and an integrated luminosity of
      100~fb$^{-1}$, 
after baseline selection, and after an additional $p_{\perp}>200$~GeV
cut (last column). The resonance width over its mass is $\Gamma/m\simeq$~10\%.}
    \label{yields1}
  \end{center}
\end{table}
In the calculations shown in Table~\ref{yields1},
in order to realistically estimate the heavy resonance parameter
space that can be probed at LHC as a function of luminosity, we
have normalized the efficiencies to the published
results by the LHC experiments. In this way, we
maintain realistic non-resonant background levels and to some
extent take systematic differences 
between the full
simulation used by the experiments and the fast simulation used here
partially into account. 
The signal efficiency $\times$ acceptance
is at the level of few \% and
based on the signal cross-section expectations summarized in
Fig.~\ref{xsectionsVsgV}, observable excesses in $M_{bb}$
should be expected for broad resonance masses below 2~TeV,
even for present LHC luminosities.

\begin{figure}[hbt!]
\begin{center}
\resizebox{0.5\textwidth}{!}{
\includegraphics{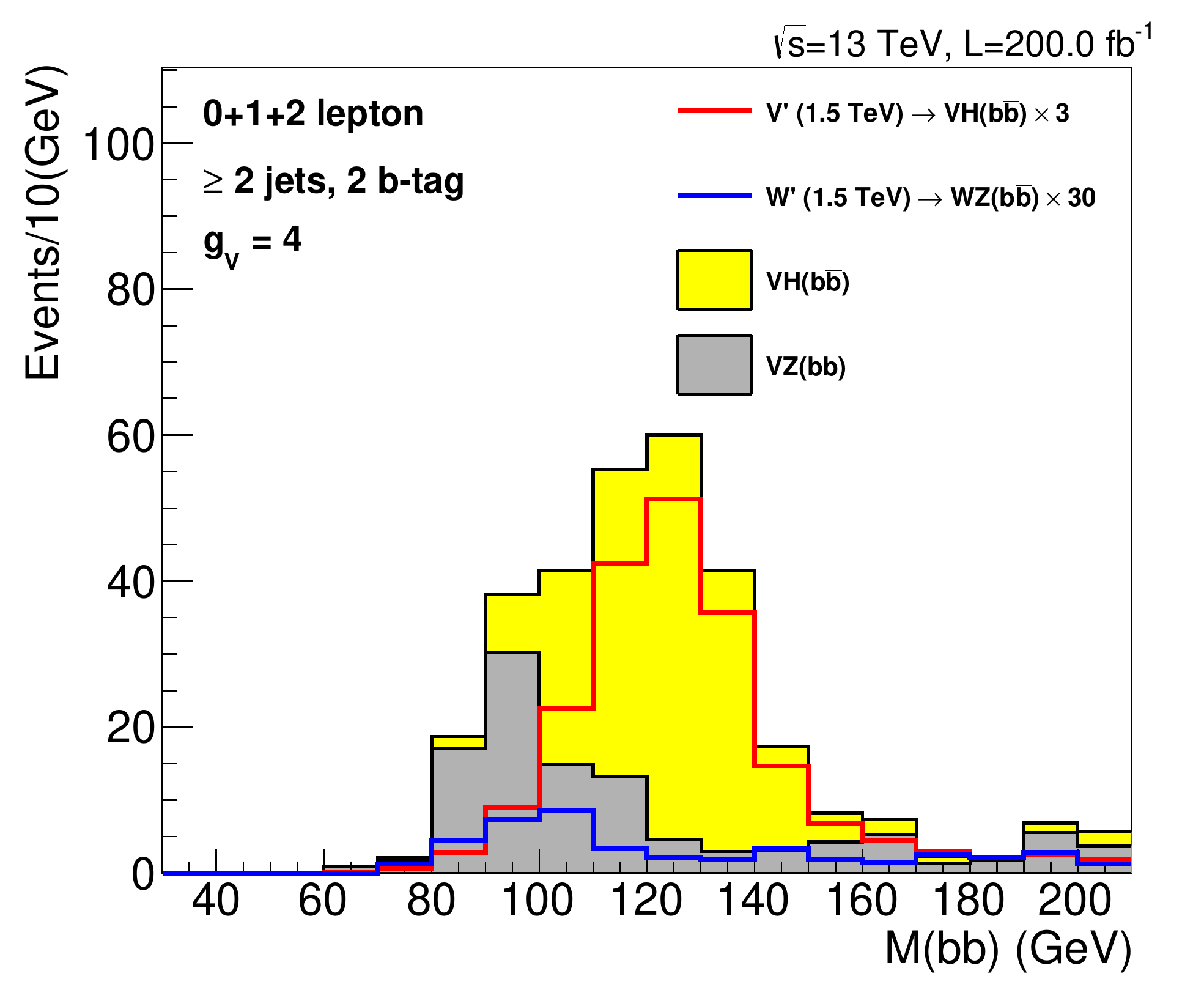}
}
\caption{Reconstructed bottom anti-bottom quark invariant mass for
SM $VH\rightarrow \ell(\nu)\ell(\nu)b\bar{b}$ and $VZ\rightarrow \ell(\nu)\ell(\nu)b\bar{b}$ 
production (stacked yellow and gray histograms, respectively), 
for an integrated luminosity of 200~fb$^{-1}$. 
The HVT model predictions for $M_{V^\prime}=1.5$~TeV and $g_V=4$, with
$VH$ shown in red and $WH$ in blue, are overlaid.
The plot includes all final states with 0, 1 and 2 leptons.
A $p_{\perp}>300$~GeV cut has been applied to the $b\bar{b}$ system. 
}
\label{Mbbmodel}
\end{center}
\vspace{-5mm}
\end{figure}

\begin{figure}[hbt!]
\begin{center}
\resizebox{0.5\textwidth}{!}{
\includegraphics{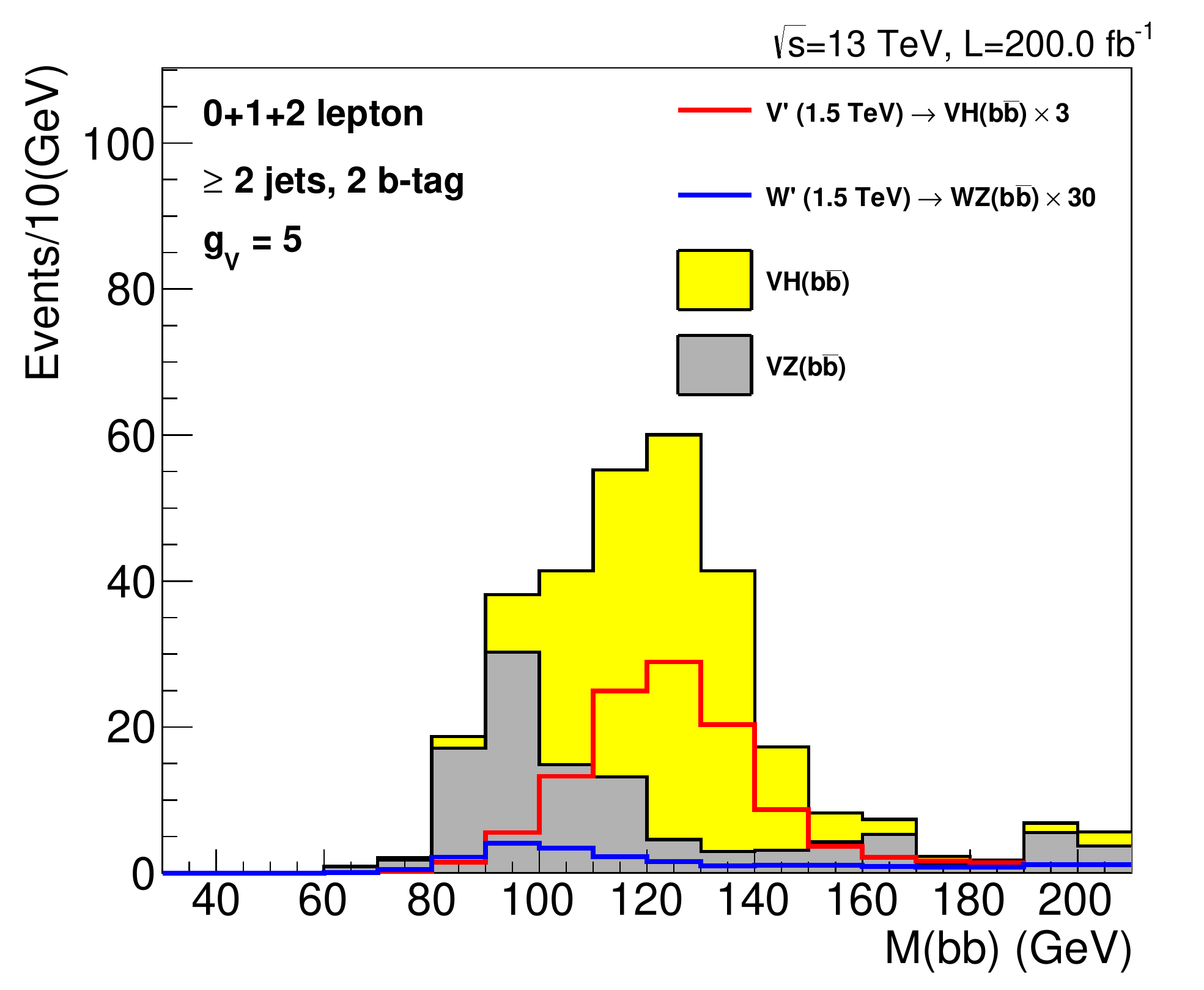}
}
\caption{Reconstructed bottom anti-bottom quark invariant mass for
SM $VH\rightarrow \ell(\nu)\ell(\nu)b\bar{b}$ and $VZ\rightarrow \ell(\nu)\ell(\nu)b\bar{b}$ 
production (stacked yellow and gray histograms, respectively), 
for an integrated luminosity of 200~fb$^{-1}$. 
The HVT model predictions for $M_{V^\prime}=1.5$~TeV and $g_V=5$, with
$VH$ shown in red and $WH$ in blue, are overlaid.
The plot includes all final states with 0, 1 and 2 leptons.
A $p_{\perp}>300$~GeV cut has been applied to the $b\bar{b}$ system. 
}
\label{Mbbmodel5}
\end{center}
\vspace{-5mm}
\end{figure}

Example scenarios with parameters $M_{Z^\prime}=1.5$~TeV,
$g_V=4$ and $g_V=5$ after baseline selection and 
a tight transverse momentum cut $p_{\perp}>300$~GeV, 
for which the bosonic decay branching ratio is dominant, 
are presented in Figures~\ref{Mbbmodel} and \ref{Mbbmodel5} respectively.
In these Figures, the combined $M_{bb}$ distribution is shown
for an integrated luminosity of 200~fb$^{-1}$.
Assuming that the new physics does not modify significantly the SM Higgs couplings,
the excess over the SM expectation 
will indicate
the presence of heavy
vector resonances decaying into $VH$ and $WZ$ final states. As we have already 
seen from Table~\ref{yields1}, most of the Higgs excess comes from the 1-lepton 
category, while all of the $ZW$ excess comes from the 1-lepton category 
but it is rather small.

\begin{figure}[hbt!]
\begin{center}
\resizebox{0.5\textwidth}{!}{
\includegraphics{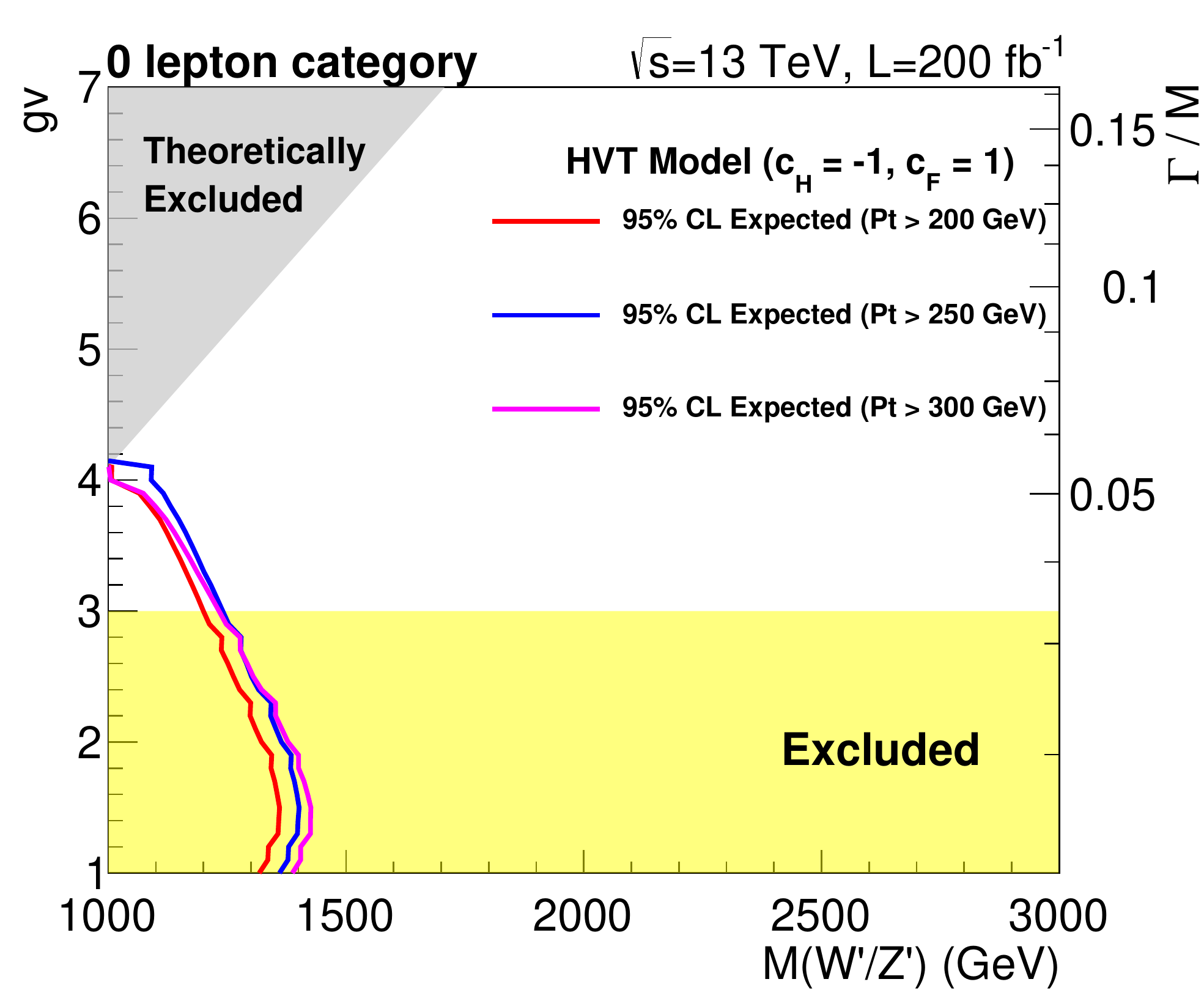}
}
\caption{Expected 95\% CL HVT limits as a function of the resonance mass and width, for an integrated luminosity of 
200~fb$^{-1}$. The limits for the 0-lepton category are shown for 
increasing $p_{\perp}>200, 250, 300$~GeV selections.
The theoretically excluded regions correspond to regions of 
small physical mass and large $g_V$ coupling, 
where the HVT model cannot reproduce the SM 
input parameters $\alpha_{EW}$, $G_F$ and $M_{Z}$.}
\label{Limits0l}
\end{center}
\vspace{-5mm}
\end{figure}

\begin{figure}[hbt!]
\begin{center}
\resizebox{0.5\textwidth}{!}{
\includegraphics{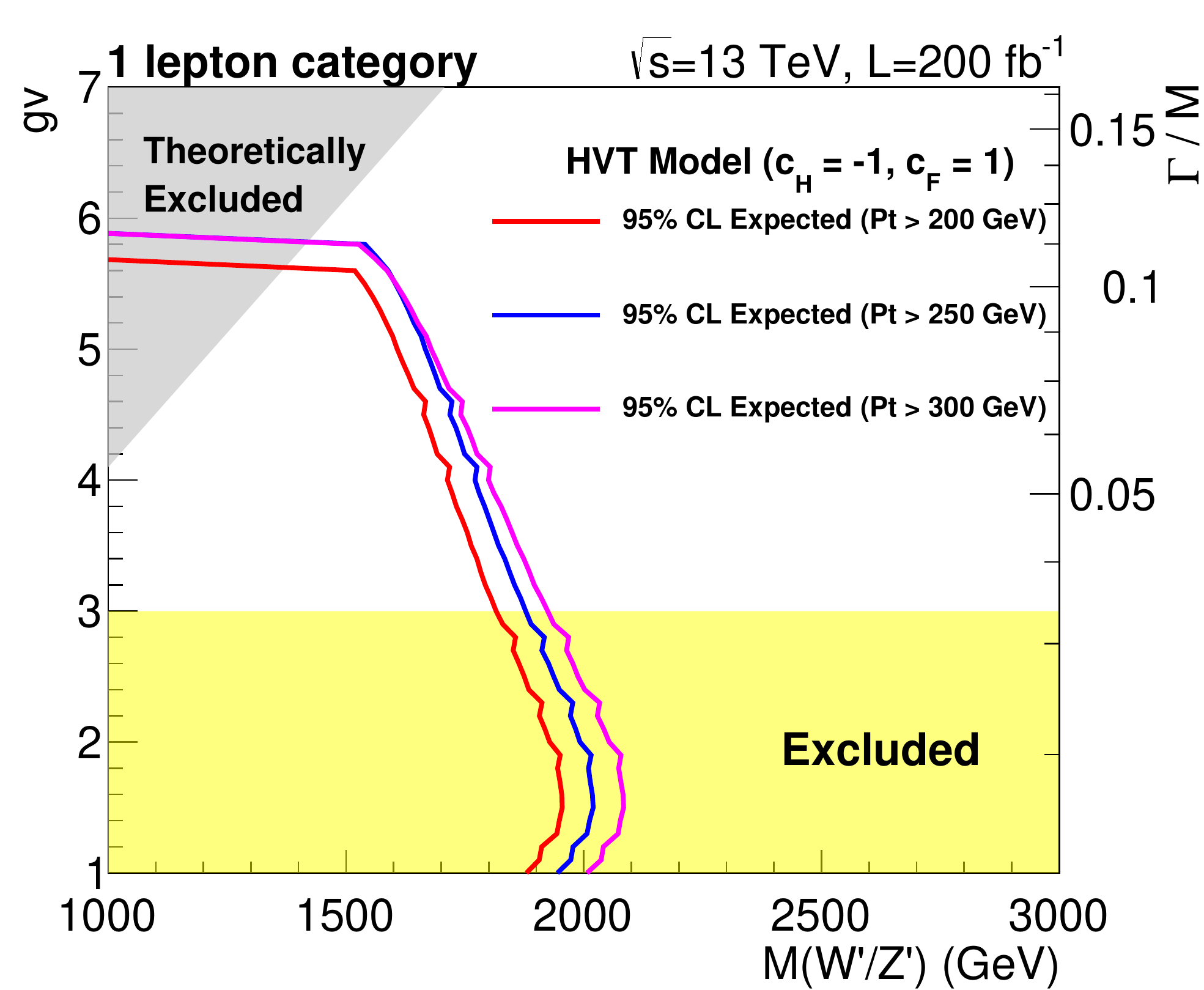}
}
\caption{Expected 95\% CL HVT limits as a function of the resonance mass and width, for an integrated luminosity of 
200~fb$^{-1}$. The limits for the 1-lepton category are shown for 
increasing $p_{\perp}>200, 250, 300$~GeV selections.
The theoretically excluded regions correspond to regions of 
small physical mass and large $g_V$ coupling, 
where the HVT model cannot reproduce the SM 
input parameters $\alpha_{EW}$, $G_F$ and $M_{Z}$.}
\label{Limits1l}
\end{center}
\vspace{-5mm}
\end{figure}

\begin{figure}[hbt!]
\begin{center}
\resizebox{0.5\textwidth}{!}{
\includegraphics{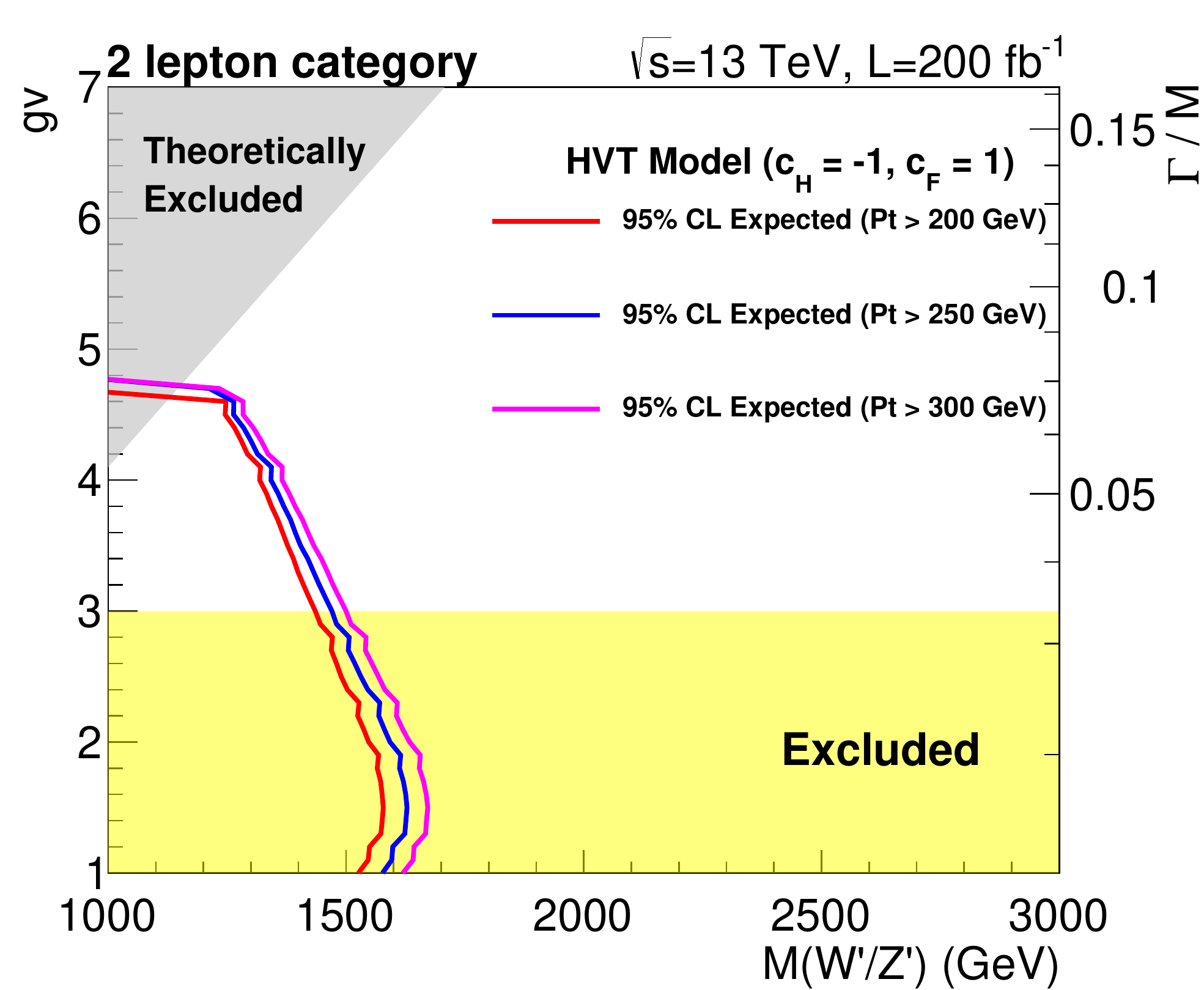}
}
\caption{Expected 95\% CL HVT limits as a function of the resonance mass and width, for an integrated luminosity of 
200~fb$^{-1}$. The limits for the 2-lepton category are shown for 
increasing $p_{\perp}>200, 250, 300$~GeV selections.
The theoretically excluded regions correspond to regions of 
small physical mass and large $g_V$ coupling, 
where the HVT model cannot reproduce the SM 
input parameters $\alpha_{EW}$, $G_F$ and $M_{Z}$.}
\label{Limits2l}
\end{center}
\vspace{-5mm}
\end{figure}

The resonance search potential for a single LHC experiment and an integrated luminosity of 200~fb$^{-1}$, 
is shown in Figures~\ref{Limits0l},~\ref{Limits1l}, and
~\ref{Limits2l}, for the 0-lepton, 1-lepton and 2-lepton categories,
respectively. 
As seen from these results, the search is sensitive to part of the
phase space not currently probed by existing searches. For
200~fb$^{-1}$, the sensitivity extends to masses up to about 2~TeV,
and resonance widths $\Gamma/m\simeq$~10\%, although  the higher the
width the lower the resonance mass reach. For $\Gamma/m\simeq$~10\%
only a mass up to 1.6~TeV can be reached. 
The reason for the search not being sensitive to larger widths is not only due 
to the reduction of cross section with $g_V$, but also due to a loss of 
efficiency because for $V^\prime$ masses greater than 1.5~GeV, the majority of 
$b$ quark pairs fall in the same jet.
More precisely, for resonance masses of 2~TeV or higher, 
the fraction of resolved $b\bar{b}$ pairs is 5\% or less, respectively.
Here it should be stressed
that an extension of the proposed search to include fat jets could
push these limits at higher values.

%
%

\begin{figure}[hbt!]
\begin{center}
\resizebox{0.5\textwidth}{!}{
\includegraphics{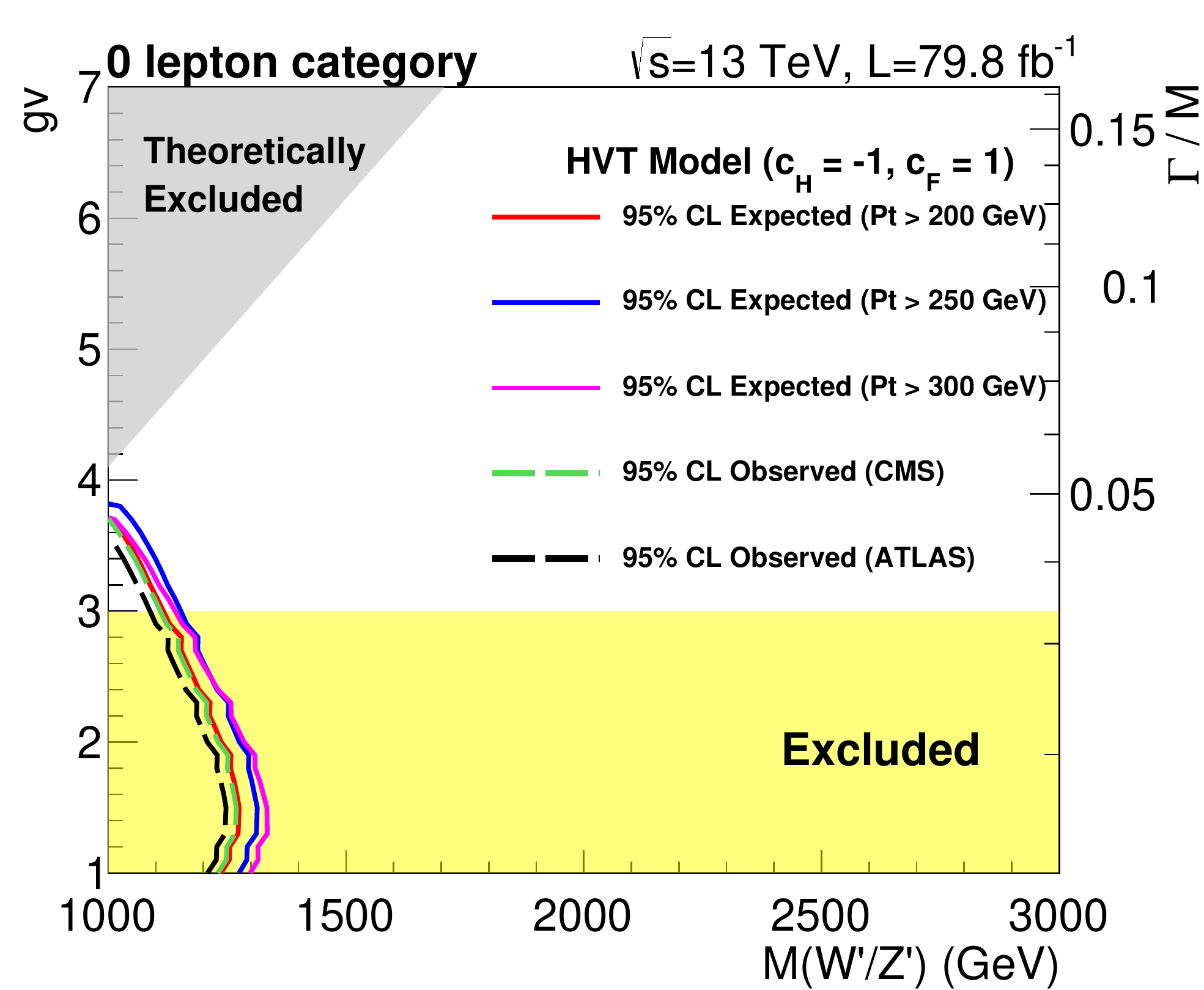}
}
\caption{Expected 95\% CL HVT limits as a function of the resonance mass and width, for an integrated luminosity of 
79.8~fb$^{-1}$. The limits for the 0-lepton category are shown for 
increasing $p_{\perp}>200, 250, 300$~GeV selections. Estimated
observed limits by ATLAS and CMS are also shown.
The theoretically excluded regions correspond to regions of 
small physical mass and large $g_V$ coupling, 
where the HVT model cannot reproduce the SM 
input parameters $\alpha_{EW}$, $G_F$ and $M_{Z}$.}
\label{Limits0ld}
\end{center}
\vspace{-5mm}
\end{figure}

\begin{figure}[hbt!]
\begin{center}
\resizebox{0.5\textwidth}{!}{
\includegraphics{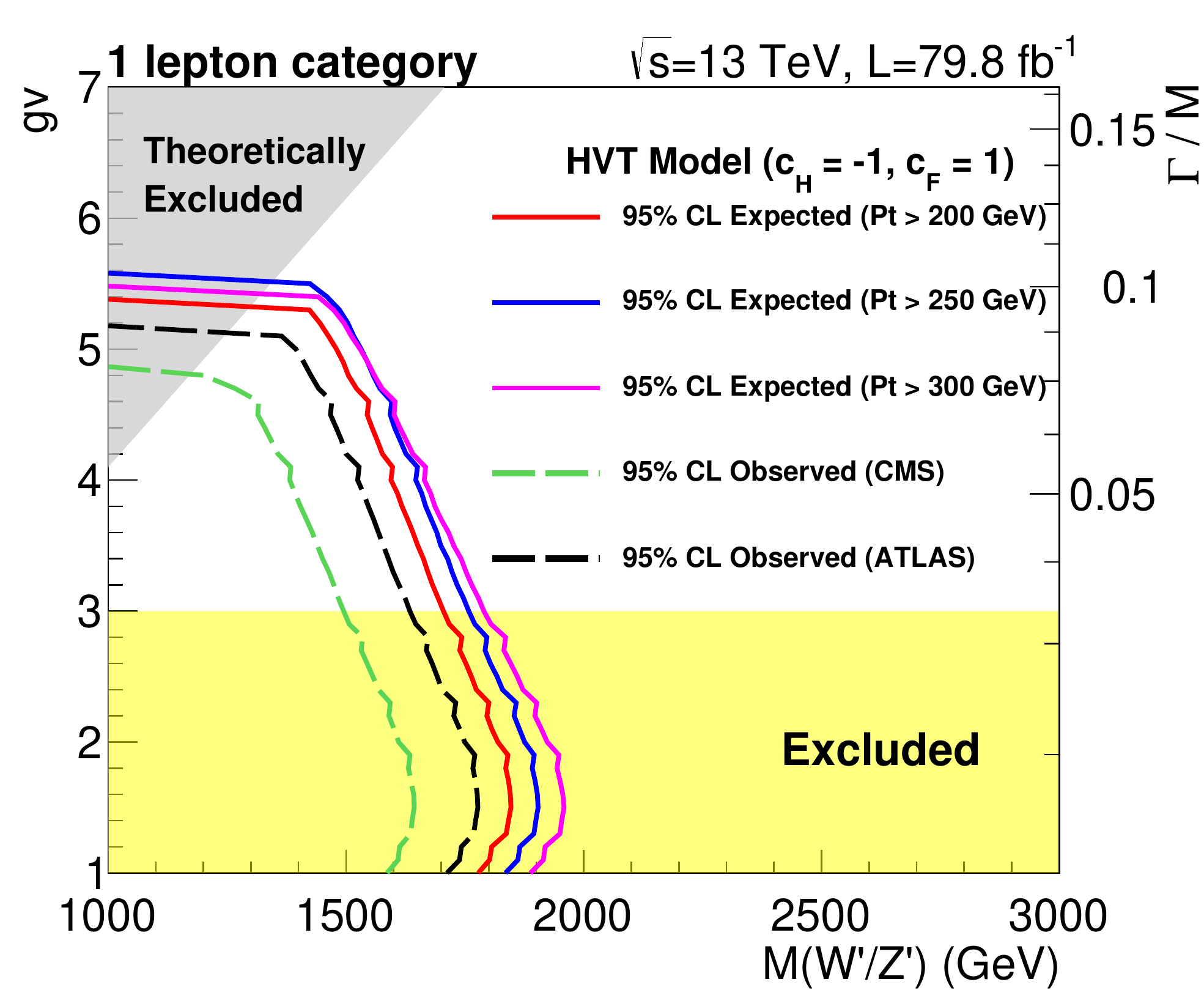}
}
\caption{Expected 95\% CL HVT limits as a function of the resonance mass and width, for an integrated luminosity of 
79.8~fb$^{-1}$. The limits for the 1-lepton category are shown for 
increasing $p_{\perp}>200, 250, 300$~GeV selections. Estimated
observed limits by ATLAS and CMS are also shown.
The theoretically excluded regions correspond to regions of 
small physical mass and large $g_V$ coupling, 
where the HVT model cannot reproduce the SM 
input parameters $\alpha_{EW}$, $G_F$ and $M_{Z}$.}
\label{Limits1ld}
\end{center}
\vspace{-5mm}
\end{figure}

\begin{figure}[hbt!]
\begin{center}
\resizebox{0.5\textwidth}{!}{
\includegraphics{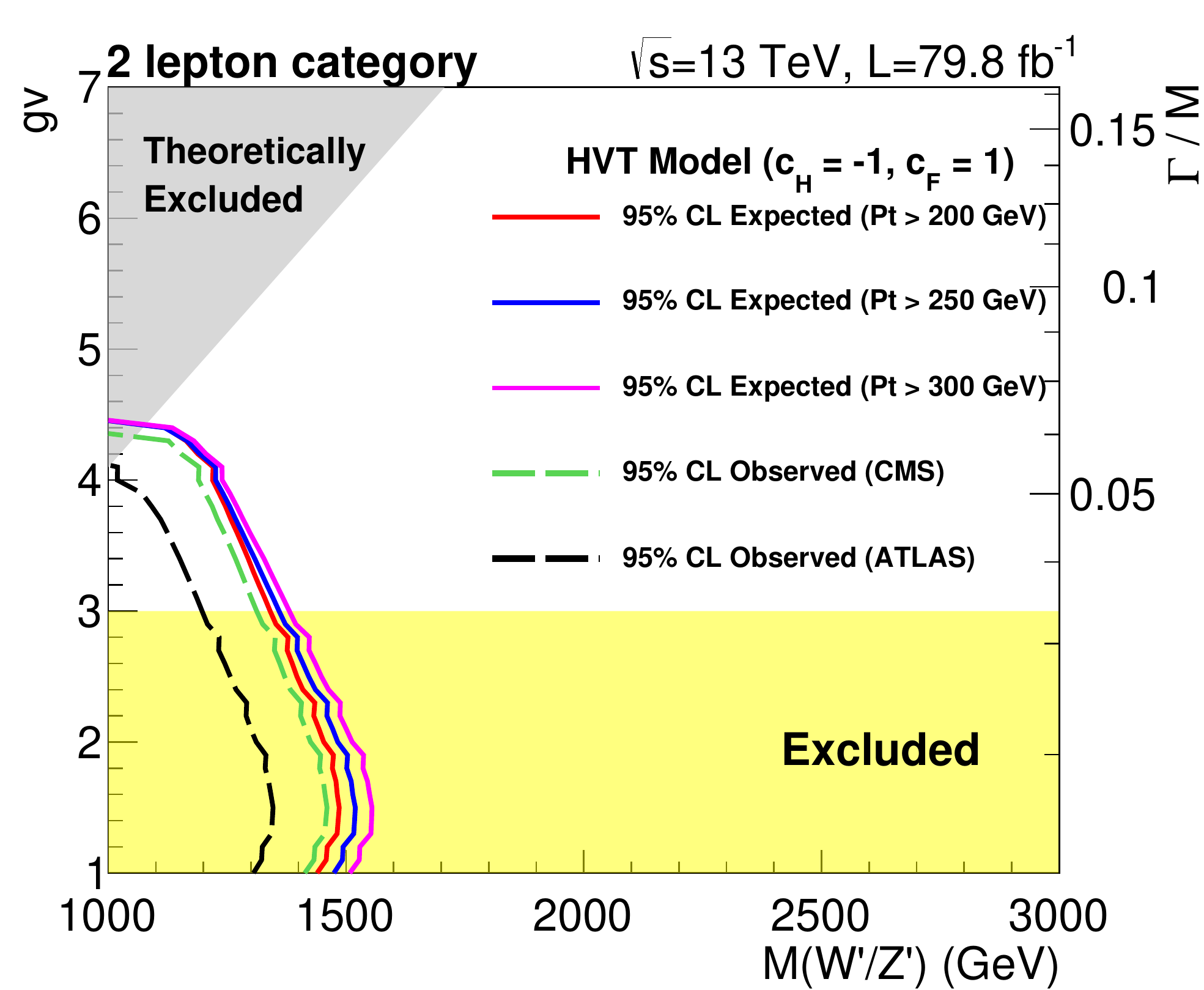}
}
\caption{Expected 95\% CL HVT limits as a function of the resonance mass and width, for an integrated luminosity of 
79.8~fb$^{-1}$. The limits for the 2-lepton category are shown for 
increasing $p_{\perp}>200, 250, 300$~GeV selections. Estimated
observed limits by ATLAS and CMS are also shown.
The theoretically excluded regions correspond to regions of 
small physical mass and large $g_V$ coupling, 
where the HVT model cannot reproduce the SM 
input parameters $\alpha_{EW}$, $G_F$ and $M_{Z}$.}
\label{Limits2ld}
\end{center}
\vspace{-5mm}
\end{figure}

In an attempt to test the prediction of the model against the
$\sim$80~fb$^{-1}$ data recently reported by ATLAS and CMS, we
estimate the observed limits as shown in
Figures~\ref{Limits0ld},~\ref{Limits1ld}, and~\ref{Limits2ld}. The
observed limits are based on the signal strength values reported by
the experiments in each category. The bands corresponding to the
uncertainties in the Higgs signal strength $\mu$ are not shown. 
%
The observed limits should be compared to the expected limits with $p_{\perp}>200$~GeV.
As seen in these results, the new LHC data already exclude new parts
of the allowed resonance mass-width phase space. 
For instance from Fig.~\ref{Limits1ld}, HVT resonances with $\Gamma/m=9\%$
and mass up to 1.5~TeV, are excluded by the ATLAS experiment.
The observed limits in the 1-lepton and 2-lepton categories exclude
lower masses and widths than expected because of the corresponding
excesses observed by the experiments. However, the uncertainties in
each category are still large and more data are needed that will first
settle if there is indeed an excess, and then allow to examine the
compatibility of the pattern of the excess with the HVT model.
It should be pointed out that the signal cross-sections were
calculated at leading order, thus an additional theory uncertainty
due to missing higher order corrections should be considered. These theoretical 
uncertainties are expected to be small with respect to the rest of
the systematic uncertainties of this analysis.

%
%
%

%
%
%
%
%
%
%

%
%

\section{Summary and Conclusions}
In this work we explored the potential of a novel search of heavy vector resonances 
of non-zero natural width, decaying almost exclusively in
the $H\rightarrow b\bar{b}$ and $Z\rightarrow b\bar{b}$ final states
in association with a leptonically decaying $V$ ($Z$ or $W$)
and $W$-only, respectively.
For large $g_V$ coupling of the exotic resonance 
to the $W$ and $Z$ bosons and the SM Higgs,
the branching ratio to
$VH$ and to two gauge bosons $VV$ dominates, leading to
a simultaneous excess to both Higgs $WH\rightarrow \ell\nu b\bar{b}$, $ZH\rightarrow \ell\ell b\bar{b}$ and
non-Higgs $WZ\rightarrow \ell\nu b\bar{b}$ final states.

We showed that excesses of varying strengths should be observed in all final
states (0, 1 or 2 leptons).
The fact that the relative strengths of these excesses depend on branching ratios
and efficiencies, provides a clear signature of the presence of heavy resonances or even their
low mass tails. 
For a luminosity accessible by the LHC experiments of 200~fb$^{-1}$, the search is sensitive 
up to resonance masses of 2~TeV and widths $\Gamma/m$ of 10\%.
A first test of a heavy vector triplet model against ATLAS and CMS data 
in terms of expected and observed limits was presented.
Although experimental uncertainties on the $VH\rightarrow b\bar{b}$ 
signal strength $\mu_{VH}$ are still large, the LHC experiments are expected to 
accumulate significant amounts of data in 2018 and beyond, making the search 
proposed here a very useful tool in evaluating potential excesses in Higgs 
yield measurements when the Higgs is produced in association with a weak vector boson.

\section*{Acknowledgements}
This work was  supported by the Taiwanese Ministry of Science and Technology 
under grant number 106-2112-M-002-011-MY3.




\begin{thebibliography}{}
\bibitem{Pati}
J. C. Pati and A. Salam,
Lepton Number as the Fourth Color,
Phys. Rev. D 10 (1974) 275.

\bibitem{Georgi}
H. Georgi and S. Glashow,
Unity of All Elementary Particle Forces,
Phys. Rev. Lett. 32 (1974) 438.

\bibitem{Fritzcsh}
H. Fritzsch and P. Minkowski,
Unified Interactions of Leptons and Hadrons,
Annals Phys. 93 (1975) 193.

\bibitem{Eichten}
E. Eichten and K. Lane,
Low-Scale technicolor at the Tevatron and LHC,
Phys. Lett. B 669 (2008) 235.

\bibitem{Contino}
R. Contino,
The Higgs as a Composite Nambu-Goldstone Boson,
arXiv:1005.4269 [hep-ph] (2010).

\bibitem{lHiggs1}
T. Han, H. E. Logan, B. McElrath, and L.-T. Wang,
Phenomenology of the little Higgs model,
Phys. Rev. D 67 (2003) 095004.

\bibitem{lHiggs2}
M. Schmaltz and D. Tucker-Smith,
Little Higgs Theories,
Ann. Rev. Nucl. Part. Sci. 55 (2005) 229.

\bibitem{lHiggs3}
M. Perelstein,
Little Higgs models and their phenomenology,
Prog. Part. Nucl. Phys. 58 (2007) 247.

\bibitem{Zprime1}
V. D. Barger, W.-Y. Keung, and E. Ma,
A gauge model with light $W$ and $Z$ bosons,
Phys. Rev. D 22 (1980) 727.

\bibitem{Zprime2}
E. Salvioni, G. Villadoro, and F. Zwirner,
Minimal $Z^\prime$ models: present bounds and early LHC reach,
JHEP 09 (2009) 068.

\bibitem{Wprime}
C. Grojean, E. Salvioni, and R. Torre,
A weakly constrained $W^\prime$ at the early LHC,
JHEP 07 (2011) 002.

\bibitem{ATLASVV_jjjj_19Dec17}
ATLAS Collaboration,
Search for diboson resonances with boson-tagged jets in pp collisions
at $\sqrt{s}=13$ TeV with the ATLAS detector,
Phys. Lett. B 777 (2017) 91.

\bibitem{ATLASVV_02lep_19Aug17}
ATLAS Collaboration,
Searches for heavy $ZZ$ and $ZW$ resonances
in the $\ell\ell qq$ and $\nu\nu qq$ final states in pp collisions
at $\sqrt{s}=13$ TeV with the ATLAS detector,
arXiv:1708.09638 [hep-ex] (2017).

\bibitem{ATLASVV_1lep_19Oct17}
ATLAS Collaboration,
Search for $WW/WZ$ resonance production in $\ell\nu qq$ final states in pp collisions
at $\sqrt{s}=13$ TeV with the ATLAS detector,
arXiv:1710.07235 [hep-ex] (2017).

\bibitem{ATLASVHbb17Dec17}
ATLAS Collaboration,
Search for heavy resonances decaying into a $W$ or $Z$
boson and a Higgs boson in final states with leptons and b-jets
in 36 fb$^{-1}$ of $\sqrt{s}=13$ TeV pp collisions with the ATLAS detector,
arXiv:1712.06518 [hep-ex] (2017).

\bibitem{ATLASVHbb_jj_21Jul17}
ATLAS Collaboration,
Search for heavy resonances decaying into a $W$ or $Z$
boson and a Higgs boson in the $qqbb$ final states in pp collisions
at $\sqrt{s}=13$ TeV with the ATLAS detector,
Phys. Lett. B 774 (2017) 494.

\bibitem{CMSVHbb_leptons_8Jul18}
CMS Collaboration,
Search for heavy resonances decaying into a vector boson and a
Higgs boson in final states with charged leptons, neutrinos and b quarks at $\sqrt{s}=13$
TeV,
arXiv:1807.02826 [hep-ex] (2018).

\bibitem{CMSVV_llqq_6Dec17}
CMS Collaboration,
Search for new heavy resonances decaying into a $Z$ boson and a
massive vector boson in the $2\ell 2q$ final state at $\sqrt{s}=13$
TeV,
CMS-PAS-B2G-17-013 (2017).

\bibitem{CMSVV_lvqq_12Dec17}
CMS Collaboration,
Search for heavy resonances decaying to pairs of vector bosons
in the $\ell\nu q\bar{q}$ final state at $\sqrt{s}=13$ TeV,
CMS-PAS-B2G-16-029 (2017).

\bibitem{CMSVV_vvqq_11Jul17}
CMS Collaboration,
Search for heavy resonances decaying into a $Z$ boson and a
vector boson in the $\nu\nu q\bar{q}$  final state,
CMS-PAS-B2G-17-005 (2017).

\bibitem{CMS_VH_17}
CMS Collaboration,
Search for heavy resonances that decay into a vector boson and a Higgs
boson in hadronic final states at $\sqrt{s}=13$ TeV,
Eur. Phys. J. C77 (2017) 9, 636.

\bibitem{CMSVVandVH_17}
CMS Collaboration,
Combination of searches for heavy resonances decaying to $WW$, $WZ$, $ZZ$,
$WH$, and $ZH$ boson pairs in proton~proton collisions
at $\sqrt{s}=8$ and 13 TeV,
Phys. Lett. B774, 533 (2017).

\bibitem{ATLASHeavyCombination}
ATLAS Collaboration,
Combination of searches for heavy resonances decaying into bosonic and leptonic 
final states 
using 36 fb$^{-1}$ of proton~proton collision data at $\sqrt{s}=13$ TeV with the ATLAS detector,
Phys.~Rev.~D~98 (2018)~5.

\bibitem{ourCHpaper1}
M. Hoffmann, A. Kaminska, R. Nicolaidou, S. Paganis,
Probing Compositeness with Higgs Boson Decays at the LHC,
Eur. Phys. J. C74 (2014) 11, 3181.

\bibitem{hvt1}
D. Pappadopulo, A. Thamm, R. Torre and A. Wulzer,
Heavy Vector Triplets: Bridging Theory and Data,
JHEP 09 (2014) 060.

\bibitem{ATLASSMvh}
ATLAS Collaboration,
Observation of $H\rightarrow bb$ decays and $VH$ production with the ATLAS detector,
arXiv:1808.08238 [hep-ex] (2018).

\bibitem{CMSSMvh}
CMS Collaboration,
Observation of Higgs boson decay to bottom quarks
CMS-PAS-HIG-18-016 (2018).

\bibitem{madgraph5}
J.~Alwall et. al.
Madgraph 5 : Going beyond,
arXiv:1106.0522v1 [hep-ph] (2011).

\bibitem{Pythia8:2008}
T. Sjostrand, S. Mrenna, P. Skands,
A Brief Introduction to PYTHIA 8.1,
Comput. Phys. Commun. 178 (2008) 852.

\bibitem{delphes}
S. Ovyn, X. Rouby, V. Lemaitre,
DELPHES, a framework for fast simulation of a generic collider experiment,
arXiv:0903.2225 [hep-ph] (2009).

\bibitem{asymp}
 G. Cowan, K. Cranmer, E. Gross, O. Vitells,
 Asymptotic formulae for likelihood-based tests of new physics,
 Eur. Phys. J. C71 (2011) $1-19$.

\end{thebibliography}
\end{document}